\documentclass[12pt]{iopart}

\usepackage{iopams}
\usepackage{graphicx}
\begin{document}

\title{Resource allocation pattern in infrastructure networks}

\author{Dong-Hee Kim and Adilson E. Motter}
\address{Department of Physics and Astronomy and Northwestern Institute on
Complex Systems, Northwestern University, Evanston, IL 60208, USA}
\ead{dongheekim@northwestern.edu and motter@northwestern.edu}

\begin{abstract}
Most infrastructure networks evolve and operate in a decentralized fashion,
which may adversely impact the allocation of resources across the system. 
Here we investigate this question by focusing on the relation between
capacity and load in various such networks. We find that, due to network
traffic fluctuations, real systems tend to have larger unoccupied portions
of the capacities---{\it smaller load-to-capacity ratios}---on network 
elements with {\it smaller} capacities,
which contrasts with key assumptions involved in previous studies. 
This finding suggests that infrastructure networks have evolved
to minimize local failures but not necessarily large-scale failures that
can be caused by the cascading spread of local damage.
\end{abstract}


\vspace{2pc}

\vspace{28pt plus 10pt minus 18pt}
\noindent{\small\rm Published in: 
{\it J. Phys. A: Math. Theor. 41 (2008) 224019.}\par}

\maketitle

\section{Introduction}
Power outages, Internet congestion, traffic jams, and many other problems 
of social and economical interest are ultimately limited by the physical
assignment of resources in infrastructure networks. 
The recent realization that numerous such systems can be modeled 
within the common framework of complex networks~\cite{review} 
has stimulated several theoretical studies on network resilience 
\cite{Albert2000,Callaway2000,Cohen2000,Shargel2003,Gallos2005,Arenas2001,
Toroczkai2004,Noh2005,Sreenivasan2007,Holme2002,Nishikawa2002,Moreno2003,
Watts2002,DHKim2005,Motter2002,Crucitti2004,EJLee2005,Schafer2006,Motter2004}.
However, despite much advance~\cite{Barrat2004}, the relation between the
large-scale allocation and actual usage of resources in distributed
infrastructure systems is a question that goes beyond previous complex
network research.
Here we propose to cast this question as a statistical physics problem 
by exploring that the dynamics of many such systems can be modeled 
as a network transport process.
For example, website browsing and e-mail communication are based on
packet transport through the Internet; 
movement of people and goods is heavily based on road, rail, and 
air transportation networks;
public utility services depend on the transport of energy, water and 
gas through supply networks. In these examples, the transport of
packets, passengers, and physical quantities creates traffic loads that
must be handled by nodes and links of the underlying networks.  
Because the capacities  of nodes and links are limited by cost and
availability of resources, the proper allocation of capacities is 
an essential condition for the robust and cost-effective operation of 
infrastructure networks.

\begin{figure}
\center{
\includegraphics[width=0.7\textwidth]{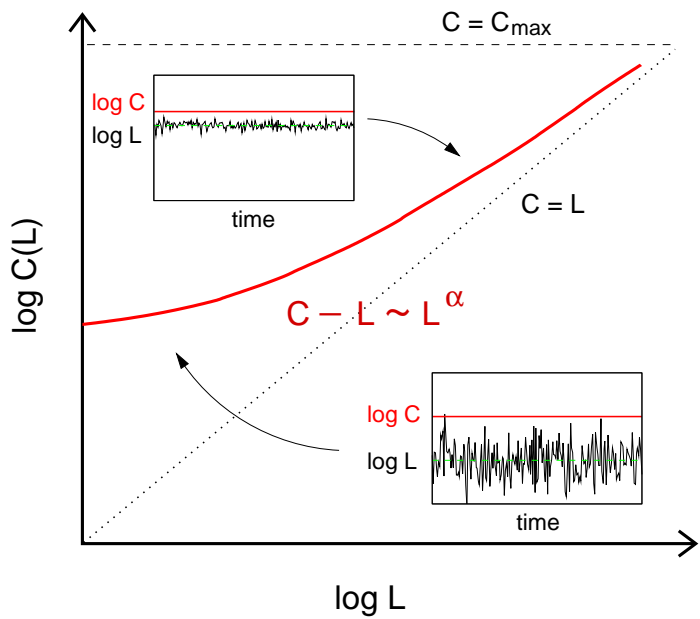}
\center}
\caption{
\label{fig1}
Typical capacity--load relation observed in real infrastructure networks. 
The nonlinear behavior of the capacity contrasts with the 
linear models 
$C\propto L$ hypothesized in previous network-theoretical work. 
Insets indicate traffic fluctuations as the expected origin 
of the observed nonlinearity.
}
\end{figure}

In this paper, we investigate this question by focusing on the 
relationship between {\it capacity} and {\it load} from the perspective of a 
decentralized  optimization between {\it robustness} and {\it cost}. 
By analyzing four types of infrastructure networks, the air transportation, 
highway, power-grid and Internet router network, we find empirically that 
the capacity--load relation is mainly determined by the relative importance
given to the cost: if robustness is much more important than cost,
the capacity $C$ approaches the {\it line of maximum robustness}
$C=C_{\max}$, where $C_{\max}$ is the maximum available capacity,
irrespective of the load; if cost is a strongly limiting factor,
the capacity approaches the {\it line of maximum efficiency} $C=L$, for
all values of the load $L$.
The real systems analyzed fall in between these two extremes and exhibit an
unanticipated nonlinear behavior, which, as shown schematically in \Fref{fig1},
is very different from the constant~\cite{Holme2002,Moreno2003}, 
random~\cite{Watts2002,DHKim2005}, and linear~\cite{Motter2002,Crucitti2004,EJLee2005,Schafer2006}
assignments of capacities considered in previous models.
We study this nonlinearity using the concept of {\it unoccupied capacity}, 
the difference between the capacity and the time-averaged load.
It follows, surprisingly, that the percentage of unoccupied capacity 
is smaller for network elements with larger capacities. 
Interpreting this as a result of a decentralized evolution in which 
capacities and loads are allocated or reallocated in response to increasing 
load demand~\cite{Dobson2007}, 
we demonstrate the observed behavior using a traffic model
devised to minimize the probability of overloads in a scenario of 
fluctuating traffic and limited availability of resources.
Our model shows that the reduction of the unoccupied capacity
is a consequence of the reduction of the traffic fluctuations on 
highly loaded elements, but it also shows that the probability of 
overloads can be larger on elements with {\it larger} capacities.

\section{Empirical Capacity-Load Characteristics}
We consider four different types of real-world infrastructure networks:
{\it air transportation network}, using seat occupation data of aircraft
operating between $1449$ US and foreign airports available at the Bureau of 
Transportation Statistics database (http://www.bts.gov);
{\it highway network}, using traffic data of the state of Colorado for $1559$
highway segments available at the Colorado 
Department of Transportation database (http://www.dot.state.co.us);
{\it power-grid network}, using available data for $5885$ transmission
lines of the Electric Reliability Council of Texas (http://www.ercot.com);
{\it Internet router network}, using packet traffic data measured
by the Multi Router Traffic Grapher (http://oss.oetiker.ch/mrtg/) on  
$721$ routers of the ABILENE backbone, MIT and Princeton University.

\begin{figure}
\center{
\includegraphics[width=1.0\textwidth]{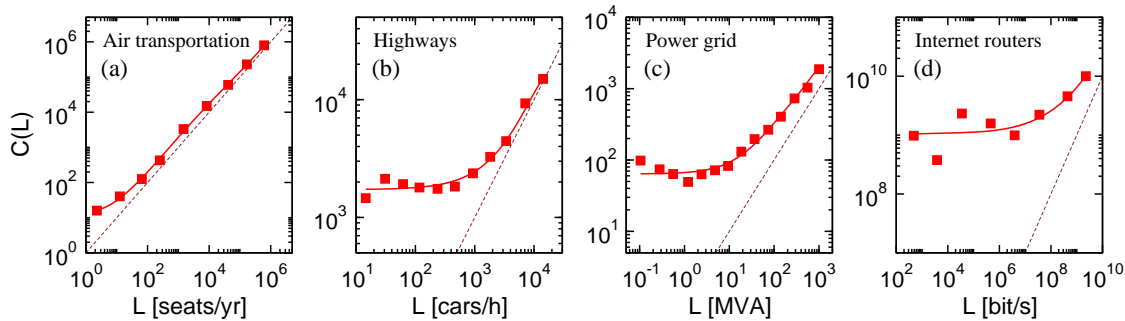}
\center}
\caption{
\label{fig2}
Capacity--load relation of four real networks.
(a) Total number of occupied ($L$) versus available
seats ($C$) in aircraft departing from and arriving at
US and international airports in $2005$.
(b) Design hourly volume ($L$) versus
estimated capacity ($C$) of Colorado highway segments in $2005$.
(c) Apparent power ($L$) versus corresponding capacity ($C$) of
transmission lines in the power grid of Texas at the summer peak 
in $2000$.
(d) Monthly averaged traffic ($L$) versus bandwidth ($C$) of the
router interfaces of the ABILENE backbone, MIT and Princeton University networks
in June $2006$. The filled boxes with curve fits indicate the averaged 
capacity--load relation $C(L)$ calculated in a logarithmic scale. 
The line of maximum efficiency $C=L$ (dashed line) is shown for comparison with the real data.
}
\end{figure}

\Fref{fig2} shows the relation between the time-averaged load and
capacity of the network elements in a log-binned scale. The
air transportation network [\Fref{fig2}(a)] is very close to the line of
maximum efficiency $C=L$, indicating effective allocation of resources, 
which is likely to be a consequence of the high costs of
air transportation combined with flexibility to
adjust seat availability to demand. The highway network shows
less efficient behavior in the region of small loads [\Fref{fig2}(b)], 
a feature that may provide alternative routes for congested traffic.\footnote{Naturally, not all unoccupied capacities correspond to alternative routes.  The extent to which they can be used to alleviate congestion and overloads in neighboring nodes and links is an interesting open problem for future research. As popularized in Ref. \cite{selfish}, decentralized routing generally produces suboptimal load distributions, which combined with the unoccupied capacities may lead to novel approaches to prevent overloads \cite{Motter2004}.}
A similar pattern is observed in the power-grid network [\Fref{fig2}(c)] 
although, compared to the highway network, the power grid has larger unoccupied 
capacities for the heavily loaded components of the network. These
unoccupied capacities can be useful
for the dispatch of power generation to adjust to
specific market, weather, and demand conditions. The Internet router network 
[\Fref{fig2}(d)] shows
weaker dependence of the capacity on the load
than those found in the other networks, which is
probably 
due to the discreteness of the commercially available router bandwidths, 
the fast growing bandwidth demand, and the tendency 
to simultaneously upgrade groups of routers 
regardless of their individual loads. Therefore, while the capacity--load 
relation depends on the specific network, the pattern of this dependence can be
understood in connection with a trade-off between the robustness and the cost of
capacities in the construction and maintenance of the system.
 
For quantitative characterization, we define the 
{\it efficiency coefficient} $\varepsilon$ of a network 
as the ratio between the total load and total capacity,
\begin{equation}
\varepsilon = \frac{\sum_i L_i}{\sum_i C_i},
\end{equation}
where the sums are taken over all components of the system.
This quantity provides a measure of the importance of the cost. 
As the cost becomes more important, the capacity is expected to 
approach the load in order to prevent overallocation of resources, 
which increases $\varepsilon$; 
when the robustness is more important, the capacity is expected to be 
much larger than the load,
which decreases $\varepsilon$. 
We find that the efficiency coefficient is $\varepsilon=0.73$ for the air transportation 
network, $0.54$ for the highway network, $0.29$ for the power-grid network, 
and $0.06$ for the Internet router network.
Therefore, one can argue that 
cost is increasingly deprioritized over robustness as one goes from the air 
transportation to the highway, power-grid and  Internet router network.
In particular, the average
unused capacity reaches $94\%$ in the Internet router network as
opposed to  $27\%$ in the air transportation system.

In interpreting these results one should notice that the capacities of the air transportation network can be easily downgraded while the same does not hold true for the other networks. Power transmission lines, highways, and Internet hardware involve permanent allocation of physical capacities that cannot be dynamically adjusted or redistributed across the system. The presence of network elements (nodes and links) with finite minimum physical capacities is likely to be a contributing factor for the plateaus observed in the region of small loads [\Fref{fig2}(b)-(d)].

\section{Capacity Optimization Model}
We now analyze our empirical findings using a model based on the optimization
of capacities at the level of individual network elements.
We define a simple objective function $F_i=(1-w) R_i(C_i) + w S_i (C_i)$ for node $i$,
which incorporates competing robustness ($R_i$) and cost $(S_i)$ functions, and
where $w \in [0,1]$ represents the importance of the cost. Given functions $R_i$ and $S_i$,
the minimization of $F_i$ will lead to an optimized capacity $C_i$ for node $i$ subjected 
to the time-averaged load $L_i$, which defines a capacity--load relation $C(L)$.
To formulate this model, we consider a time-dependent transport process in which 
traffic moves from source to destination along predetermined paths.\footnote{For simplicity, 
multiple paths are not taken into account here.} 
This process includes as a special case the directed flow model
where traffic moves along the shortest paths \cite{Menezes2004a},
and it leads to a general yet mathematically treatable model that
is not dependent on the details of the network structure and routing scheme.

Within this model, we identify the time fluctuation of traffic as 
the main perturbation that can cause accidental overloading failures.
Defining the robustness function $R_i$ as the overloading probability $\xi_i$,
the objective function can be written as
\begin{equation}
\label{eq:objfn}
F_i = (1-w)\xi_i(C_i) + w \frac{C_i}{C_{\max}},
\end{equation}
where, for concreteness, we have chosen the cost $S_i$ to be
a linear function of the capacity.

To determine the overloading probability, we calculate the load 
$l_i(t)=\sum_{j,k}z_{jk;i}x_{jk;i}(t)$ on node $i$ at time $t$,
where $x_{jk;i}(t)$ is the amount of the traffic towards node $k$ 
originating from node $j$ at time $t-\tau_{jk;i}$.
Here, $z_{jk;i}$ is $1$ if $i$ lies on the path
from $j$ to $k$ and $0$ otherwise. 
If we take a time window
$\Delta t$ much larger than the autocorrelation time of 
$x_{jk;i}$'s, we can rewrite the time-averaged load $L_i$ as 
\begin{equation}
\label{eq:3}
L_i = \frac{1}{\Delta t} \int_t^{t+\Delta t} l_i(t^\prime) dt^\prime
\simeq \sum_{j,k} z_{jk;i} \langle x_{jk;i} \rangle,
\end{equation}
where $\langle \cdot \rangle$ is used for ensemble averages.
Then, given the distribution of $x_{jk;i}$, 
and thereby the load distribution $P_i(l_i)$, 
we can write the overloading probability $\xi_i$ as
\begin{equation}
\label{eq:xi}
\xi_i(C_i)=\textrm{Prob}[l_i > C_i] = \int^\infty_{C_i} P_i(l_i) dl_i
\end{equation}
for given capacity $C_i$ such that $L_i \le C_i \le C_{\max}$.
The capacity $C$ is assumed to be physically upper-bounded 
by $C_{\max}$ and lower-bounded by $L$.

For the explicit calculation of the capacity--load relation, we consider
uncorrelated and synchronized traffic fluctuations.
We consider both types of fluctuations since 
random internal fluctuations can be strongly modulated by 
external driving forces~\cite{Menezes2004b}. 
In the Internet backbone, for example, it has been observed that the
traffic dynamics is well-characterized by a Poisson process for millisecond
time scales, while long-range correlations appear for longer time scales~\cite{Karagiannis2004}.
In the systems we consider, 
synchronized fluctuations can be generally triggered by exogenous factors, 
such as weather and seasonal conditions or collective human behavior. 

{\bf Uncorrelated Fluctuations.} We consider fluctuations in which 
the traffic $x_{jk;i}$ is completely uncorrelated with the 
traffic between different source-destination nodes. 
In this regime, the quantity $x_{jk;i}$ 
can be regarded as an independent identically distributed random variable 
$r$ following a probability distribution $p(r)$.  
Assuming that $p(r)$ 
has finite moments, including average $\bar{r}$ 
and variance $s^2$, we apply the central limit theorem
to obtain a Gaussian distribution of loads,
\begin{equation}
\label{eq:pdf_uncorrel}
P_i(l_i) \simeq \frac{1}{\sigma_i\sqrt{2\pi}}
\exp\left[-\frac{(l_i-L_i)^2}{2\sigma_i^2}\right],
\end{equation}
with  average 
$L_i = \bar{r} \sum_{j,k}z_{jk;i} \equiv \bar{r} z_i$ 
and variance $\sigma_i^2 = s^2 z_i$.
The relation $\sigma_i\sim L_i^{1/2}$, a corollary of 
Eq.~(\ref{eq:pdf_uncorrel}), is in agreement with the empirical results
of previous studies \cite{Menezes2004a,Menezes2004b,Duch2006}.    
Now, using Eq.~(\ref{eq:pdf_uncorrel}) in the minimization of $F_i$ 
in Eq. (\ref{eq:objfn}), we obtain 
the capacity--load relation as $C(L)=\min\{C^\prime(L),C_{\max}\}$ with 
\begin{equation}
C^\prime(L) = \left\{ \begin{array}{ll}
L + g L^\frac{1}{2}\sqrt{\log\Omega(L)} & \textrm{if $L<L_w$}\\
L & \textrm{if $L>L_w$}, \end{array} \right .
\end{equation}
where 
$\Omega(L)\equiv\frac{1}{g\sqrt{\pi}}\frac{1-w}{w}\frac{C_{\max}}{L^{1/2}}$,
parameter $g$ denotes $\sqrt{2s^2/\bar{r}}$,
and $L_w$ satisfies $\Omega(L_w)=1$.
  
{\bf Synchronized Fluctuations.}
Since the modulation of traffic that we describe by 
synchronized fluctuations occurs in a longer time scale
\cite{Menezes2004b,Duch2006}, we neglect the travel time $\tau_{jk;i}$ 
to express $x_{jk;i}(t)\equiv x(t)$ and the synchronized traffic load 
as $l_i(t) = x(t)z_i$. 
Assuming statistical independence of $x(t)$
in different modulation periods, we use the peak value $r$ of $x(t)$ 
in each modulation period  as a reference for capacity determination.
Given the distribution $p(r)$,
we can write the overloading probability $\xi_i$ as
\begin{equation}
\xi_i (C_i) 
= \int^\infty_{C_i} P(l_i) dl_i 
= \int^\infty_{\bar{r} C_i/L_i } p(r) dr,
\end{equation}
and determine $C(L)$ by minimizing $F_i$. 
The resulting optimized capacity is $C(L)=\min\{C^\prime(L),C_{\max}\}$, with
\begin{equation}
\label{eq:c_sync}
C^\prime(L) = \left\{ \begin{array}{ll}
\frac{L}{\bar{r}} \ 
q\big(\frac{w}{1-w}\frac{L}{\bar{r} C_{\max}}\big) & 
\textrm{if $L<L_w$}\\
L & \textrm{if $L>L_w$}, \end{array} \right .
\end{equation}
where $L_w = \bar{r} C_{\max}\frac{1-w}{w}\max_r p(r)$ and 
$q(y)=r$ is obtained by inverting $y=p(r)$. 
For $y=p(r)$ having more than one solution, 
we conventionally select $q(y)$ that gives the largest capacity. 
Because we have defined $r$ as 
the maximum traffic amount of many individual traffic events 
in a modulation period, 
extreme value distributions can be used as an input for $p(r)$.
Here we numerically calculate $C(L)$ for the Gumbel distribution 
$p_g(r)=\frac{1}{\beta} \exp[-\frac{r-\mu}{\beta}-e^{-\frac{r-\mu}{\beta}}]$ 
and the Fr\'echet distribution 
$p_f(r)= \frac{\gamma}{\alpha^{-\gamma}}r^{-\gamma-1}
\exp[-\big(\frac{r}{\alpha}\big)^{-\gamma}]$,
where all parameters are positive, 
referred to as the first and second asymptotes
in the extreme value statistics literature~\cite{book:extreme}. 
These two asymptotes correspond to exponential and 
power-law initial distributions, respectively. 
The third asymptote is for bounded initial distributions 
and gives similar results when the bound of the traffic $x(t)$ becomes 
large.

\begin{figure}
\center{
\includegraphics[width=0.8\textwidth]{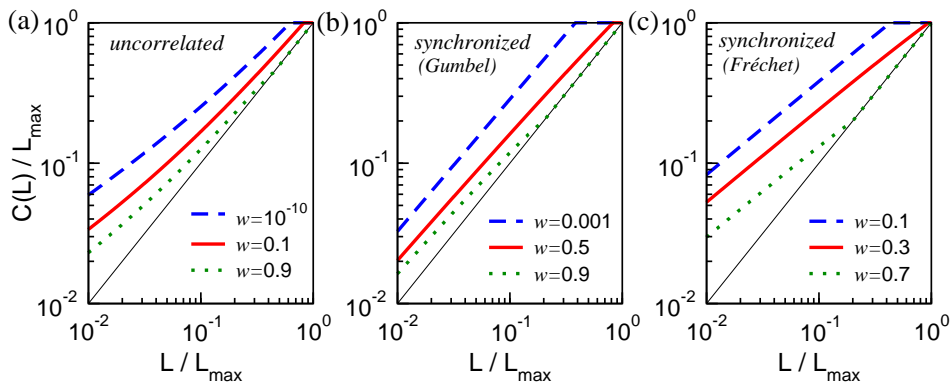}
\center}
\caption{
\label{fig3}
Capacity--load relation derived from the capacity optimization
model for three values of the weight $w$ assigned to the cost:
({\it a}) uncorrelated fluctuation regime for the model parameter $g=3$;
({\it b}) synchronized fluctuation regime for the Gumbel distribution 
with parameters $(\mu,\beta)=(100,20)$; ({\it c}) synchronized fluctuation 
regime for the Fr\'echet distribution with parameters $(\alpha,\gamma)=(1,2)$.
The capacity and load are normalized by the predefined maximum value 
$C_{\max}=L_{\max}=10^3$.
}
\end{figure}

\Fref{fig3} shows our model predictions for uncorrelated and synchronized
fluctuations. In both regimes we find that the allocation of capacities exhibits
characteristics in common with the empirical data. In particular, the calculated
$C(L)$ shows the common trend that a larger relative deviation from the
line $C=L$, representing a larger unoccupied portion of the capacity, 
is found in the region of smaller $L$. 
These results are determined by general statistical
properties of the traffic and do not depend on the details of 
the network structure and dynamics.\footnote{
However, the empirical capacity--load relation is expected to be 
partially influenced by constraints imposed by the network topology and the minimum available capacities.
}
This generality represents
an advantage over previous models based on betweenness centrality because the 
latter is only weakly correlated with the actual flows 
in the networks\footnote{For the air transportation network, whose topology is
available, the Pearson correlation coefficient between the
actual load and the betweenness centrality is 0.02.} 
and cannot be used to predict $C(L)$.  
Note that betweenness centrality only accounts for the shortest paths, while $l_i$ in Eq.~(\ref{eq:3})
accounts both for paths that are not necessarily the shortest and for non-uniform distributions of
the ``size" $x_{jk;i}$ of the individual traffic events.

\section{Discussion}
The observed nonlinearity in the capacity--load 
relation suggests that infrastructure systems have evolved under 
the pressure to minimize local failures rather than global failures. 
Previous work~\cite{Motter2004} has established that the incidence of 
large cascading failures can be reduced by shedding loads on low-load nodes,
despite the fact that this causes a concurrent increase in the incidence of
small failures. In the present model this would correspond to a higher 
probability of overloads for network elements subjected to smaller loads, 
which is the {\it opposite} of the trend observed in this study. 
Indeed, as a result of the optimization of capacities, 
the overloading probability $\xi(L)$ is an increasing function of $L$ and
differs, in particular, from capacity allocations that 
assume the same overloading probability for all the nodes.
The predicted vulnerability to large-scale failures is consistent with 
the absence of global optimization given that real infrastructure networks 
evolve in a decentralized way. In the case of the power grid, for example, 
it has been proposed~\cite{Dobson2007} that the evolution of the system 
is driven by the opposing forces of slow load increase
and corresponding system upgrades, keeping the system in a dynamic equilibrium
that balances the probability of outages. It is likely that a similar 
self-organization mechanism is at work in infrastructure systems in general,
which would further expand the concept of network self-organization 
\cite{netSOC} within transportation problems. 
While providing additional rationale 
for the decentralized optimization incorporated in our model, 
this view emphasizes that in infrastructure systems local robustness 
is prioritized at the expense of global robustness. 
These results are expected to enable researchers to build models 
to study network evolution and the impact of disturbances in complex 
communication and transportation systems.

\ack
This work was supported by NSF Grant DMS-0709212.

\end{document}